\documentclass[11pt,twoside]{article}

\usepackage{asp2010}

\resetcounters

\begin{document}
\title{Metal Abundances in Hot DO White Dwarfs: RE\,0503$-$289 and KPD\,0005+5106}  
\author{K. Werner,$^1$ T. Rauch,$^1$ E. Ringat,$^1$ and J. W. Kruk$^2$}  
\affil{$^1$Institute for Astronomy and Astrophysics, Kepler Center for
  Astro and Particle Physics, University of T\"ubingen, Sand~1, 72076 T\"ubingen, Germany\\
$^2$NASA Goddard Space Flight Center, Greenbelt, MD 20771, USA
}  

\begin{abstract}
The relatively high abundance of carbon in the hot DO
white dwarf RE\,0503$-$289 indicates that it is a descendant of a PG\,1159
star. This is corroborated by the recent detection of the extremely high abundances of
trans-Fe elements which stem from s-process nucleosynthesis in the precursor AGB
star, dredged up by a late He-shell flash and possibly amplified by radiative
levitation. On the other hand, the hottest known DO white dwarf, KPD\,0005+5106,
cannot have evolved from a PG\,1159 star but represents a distinct He-rich
evolutionary sequence that possibly originates from a binary white dwarf merger.
\end{abstract}

\begin{figure}[bth]
\begin{center}
\epsfxsize=0.90\textwidth \epsffile{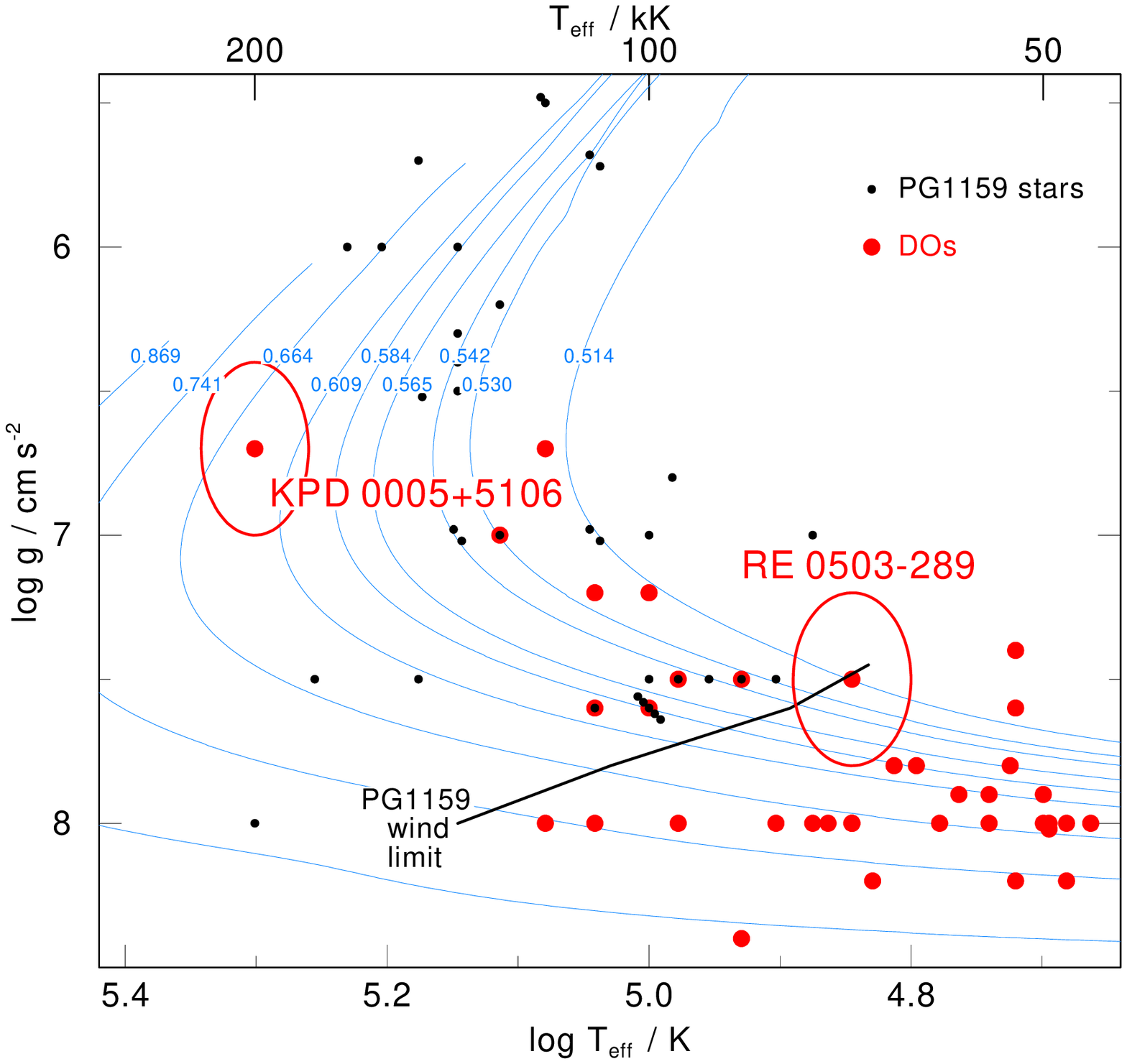}
\end{center}
\vspace{-0.5cm}
\caption{PG\,1159 stars (black small dots)  and hot DO white dwarfs (red big dots)
in the $g$--$T_{\rm eff}$--plane. The positions and error ellipses of RE\,0503$-$289 and
KPD\,0005+5106 are highlighted. Evolutionary tracks for H-deficient WDs 
\citep{althaus:2009} are labeled with the respective stellar
mass in solar units. Also shown is one of the theoretical wind limits
\citep{unglaub:2000} near which PG\,1159
stars transform into DOs because gravitational settling overcomes
radiation-driven mass loss.}\label{fig_gteff}
\end{figure}
 
\begin{figure}[bth]
\begin{center}
\epsfxsize=1.0\textwidth \epsffile{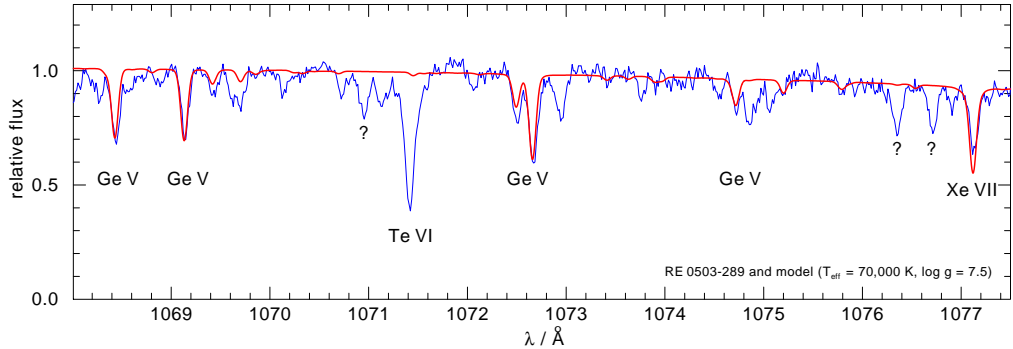}
\end{center}
\vspace{-0.5cm}
\caption{Representative detail from the FUSE spectrum of RE\,0503$-$289. Most lines
  stem from Ge\,V, one is from Xe\,VII. Note the prominent strength of the
  Te\,VI line. Many lines remain unidentified. Overplotted is the final model
  from \citet{rauch:2012}. No trans-iron group elements other than Ge, Kr, and Xe are included.}\label{fig_fuse_example}
\end{figure}

\begin{figure}[bth]
\begin{center}
\epsfxsize=0.8\textwidth \epsffile{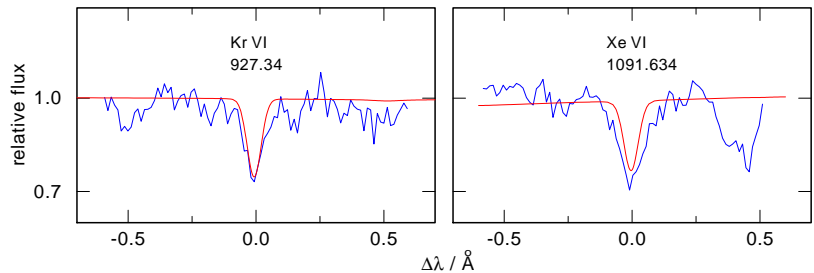}
\end{center}
\vspace{-0.5cm}
\caption{Two examples for Kr\,VI and Xe\,VI lines in RE\,0503$-$289. Overplotted
  is the final model from \citet{werner:2012}}\label{fig_krxe}
\end{figure}

\begin{figure}[bth]
\begin{center}
\epsfxsize=0.5\textwidth \epsffile{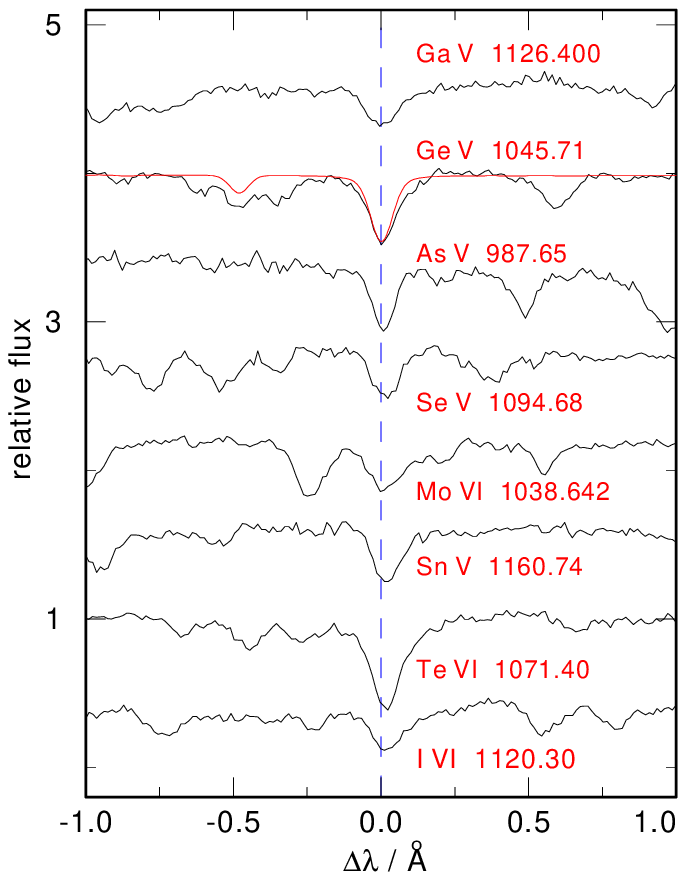}
\end{center}
\vspace{-0.5cm}
\caption{Examples of lines from eight other trans-iron elements discovered in
  RE\,0503$-$289. Only for Ge, line-profile calculations are hitherto available
  (from final model presented by
  \citealt{rauch:2012}).}\label{fig_trans_fe_lines}
\end{figure}

\section{Trans-Iron Elements in White Dwarfs}

The presence of elements beyond the iron group in white dwarfs was first proved
by \citet{vennes:2005}. They discovered Ge in three hot ($T_{\rm eff}$ =
56,000--58,000\,K) DA white dwarfs (GD\,246, Feige\,24, G191$-$B2B) with roughly
solar abundance. Subsequently, \citet{chayer:2005} detected Ge, As, Se, Sn, and
I in two cool ($T_{\rm eff}$ = 49,500\,K) DOs (HD\,149499B and HZ\,21). The
abundances are between 3 and 1000 times solar. As described in detail below,
\citet{werner:2012} discovered no less than ten trans-iron elements in the hot
DO RE\,0503$-$289. From them, Ga, Kr, Mo, and Xe are newly discovered species in
WDs. The abundances of Ge \citep{rauch:2012}, Kr, and Xe range between several
100 and several 1000 solar. In essence, extreme overabundances of trans-iron
elements are only seen in DO white dwarfs, in a temperature range of
49,500--70,000\,K.

It might be worthwhile to recapitulate that large overabundances (up to 4 dex
oversolar) of Ga, Ge, Sr, Y, Zr, Sn, and Pb where found in hot subdwarfs
($T_{\rm eff}$ = 22,000--40,000\,K;
\citealt{otoole:2004,chayer:2006,otoole:2007,naslim:2011}), although the
relevance of this fact in the context of WDs is not immediately obvious.

\section{RE\,0503$-$289}

RE\,0503$-$289 is a hot DO with $T_{\rm eff}$ = 75,000\,K, $\log g = 7.5$
(Fig.\,\ref{fig_gteff}; \citealt{barstow:1994,dreizler:1996}). It has a rather
high C abundance (e.g., \citealt{rauch:2012}), namely $\approx$3--5\% (mass
fraction). This amount is significantly larger than that found in other DOs
($<$1\%) but, on the other hand, much lower than the C abundance seen in
PG\,1159 stars (13--60\%). The ISM column density towards RE\,0503$-$289 is
extremely low ($\log n_{\rm H} =17.1$; \citealt{barstow:1994}). This uniquely
allows access to the EUV spectrum of a DO white dwarf. A remarkable spectrum was
recorded with the EUVE satellite \citep{vennes:1998}. Up to now, it was not
possible to fit that spectrum with a model atmosphere flux. It was realized that
there are unknown absorbers in the atmosphere which were not included in the
models \citep{werner:2001}.

In the years 2000 and 2001, far-UV spectra of RE\,0503$-$289 were observed with
FUSE, aiming at the ISM D/H determination along the line of sight. Since then
not a single attempt was made to analyze the photospheric spectrum. What
prevented us from analysing the FUSE data was the fact that the spectrum is
dominated by absorption lines that are not seen in any other WDs and which
remained entirely unidentified. Eventually, we realised that they stem from
highly-ionized trans-iron group elements as mentioned above (examples are shown
in Figs.\,\ref{fig_fuse_example}--\ref{fig_trans_fe_lines}).

The most serious problem for the determination of trans-iron element abundances is
the lack of atomic data. NLTE modeling is necessary and therefore oscillator
strengths are required not only for the observed line transitions but for the
entire set of lines that need to be considered in the model atoms. While for Kr
and Xe we could compile atomic data from literature \citet{werner:2012}, new
efforts are necessary for the other species. A first step was taken by new
quantum mechanical calculations for Ge\,V and VI that were immediately applied
to derive the Ge abundance in RE\,0503$-$289 \citep{rauch:2012}. A further
complication for future work is the lack of experimentally derived energy levels
so that the line positions are uncertain.

\section{KPD\,0005+5106}

Spectroscopically, KPD\,0005+5106 is classified as a hot DO
\citep{sion:1999}. However, it was shown recently that the temperature of
this star ($T_{\rm eff}$ = 200,000\,K) is much higher than previously assumed
\citep{werner:2007,wassermann:2010}. Its location in the $g$--$T_{\rm
eff}$--plane is well before the maximum-temperature ``knee'', thus, it is
strictly speaking no WD but a He-shell burning post-AGB star. The metal
abundances in the He-dominated atmosphere (mass fractions of C, N, O, Ne, Si, S,
Ca, Fe) display moderate deviations from solar abundances. They range between
0.7 and 4.3 times solar.

\section{Evolutionary Status of the Hottest DO White Dwarfs}

Figure~\ref{fig_gteff} displays the location of DO and PG\,1159 stars in the
$g$--$T_{\rm eff}$--plane. Also shown is a theoretical PG\,1159 wind limit which
means that any PG\,1159 star that approaches this line during its evolution
transforms into a DO WD. This is because gravitational settling of heavy
elements overcomes radiation-driven mass-loss in the fading stars. The location
of the wind-limit line depends on the assumed mass-loss law and is therefore
somewhat uncertain. It is probably intrinsically ``fuzzy'' because of the
metallicity dependence of the mass-loss rate. In any case, no PG\,1159 stars is
expected below that limiting line, and indeed this is supported by the
observations. On the other hand, DO WDs above this line cannot have evolved from
PG\,1159 stars. This was also concluded by \cite{quirion:2012}, who investigated
theoretically the red edge of the GW~Vir instability strip. GW~Vir stars are
pulsating PG\,1159 stars and it was originally proposed by \cite{quirion:2006}
that the red edge is a consequence of gravitational settling the removes the
driving agents, mainly C and O, from the stellar envelope. \cite{quirion:2012}
emphasize that at any time during the evolution there is no significantly
different composition in the driving region and the photosphere, so that the
location in the $g$--$T_{\rm eff}$--plane where PG\,1159 stars transform into
DOs coincides with the red edge of the GW~Vir strip. The wind limits derived by
\cite{quirion:2012} are qualitatively similar to those derived by
\cite{unglaub:2000} but, quantitatively different because of diverse assumptions.

RE\,0503$-$289 is located close to the wind limit that is displayed in
Fig.\,\ref{fig_gteff}. This fact, as well as the intermediate carbon abundance,
suggests that the star is about to transform from a PG\,1159 star into a DO WD
\citep{vennes:1998,unglaub:2000}. While C is already depleted because of
gravitational settling, the trans-iron elements are strongly enhanced by radiative
levitation. The reservoir from which the heavy metals are drawn is the former
He-rich intershell region, which dominates the envelope composition in PG\,1159
stars as the result of a late He-shell flash (e.g. \citealt{werner:06}) and
which can be enriched with s-process elements by 2--3 dex (Gallino, priv. comm;
\citealt{karakas:2007}). This circumstance could explain that hot DAs do not
show the extreme overabundances of trans-iron elements. The only disturbing fact
is that iron is not detected, suggesting that the abundance must be less than
2.5 dex subsolar \citep{barstow:2000}. This is surprising because we would
expect that Fe is also kept in the photosphere with higher abundances due to
radiative levitation. The iron abundance in PG\,1159 stars is solar
\citep{werner:2011} and diffusion calculations predict roughly solar abundances
in a DO like RE\,0503$-$289 \citep{chayer:1995}. Nickel could also pose a
problem for the interpretation of the metal abundances. The presence of Ni (0.3
dex oversolar) in HST spectra of RE\,0503$-$289 was claimed
\citep{barstow:2000}, however, from today's view this seems doubtful to us and a
reassessment of this question, including FUSE spectra,  would be useful.

DOs that have not quite reached the wind limit indicated in
Fig.\,\ref{fig_gteff} could stem from ``milder'' PG\,1159 stars in a sense, that
their C abundance was relatively low. This could result in lower mass-loss rates
and hence a shift of the wind limit towards lower gravities. One example could
be the DO PG\,0108+101 ($T_{\rm eff}$ = 95,000\,K, $\log\,g = 7.5$) which has a
C abundance similar to RE\,0503$-$289 (3\%, \citealt{dreizler:1999}). It would
be interesting to know whether this WD also exhibits large amounts of trans-iron
elements. Unfortunately, there are no FUSE observations, but HST could be used
to record a FUV spectrum.

KPD\,0005+5106 is located well before the wind limit and, thus, cannot have been
a PG\,1159 star \citep{wassermann:2010}. Its metal abundance pattern suggests a
possible relation to the RCrB stars which in turn are probably the result of a
WD merger. This evolutionary context was discussed in detail in other papers of
this conference, presented in the talks by Clayton, Reindl, and Staff.

\section{Summary and Conclusions}

Some hot DOs above the PG\,1159 wind limit are not descendants from PG\,1159
stars but belong to a distinct helium-rich post-AGB sequence (whatever its
origin is). The most prominent example is KPD\,0005+5106. RE\,0503$-$289 is a
unique white dwarf. It could be in the transition phase from the PG\,1159 into
the DO class. The high trans-iron element abundances result from s-process
enhancement, possibly amplified by radiative levitation. The two other, cooler
DOs with abundant trans-iron species mentioned in the Introduction are cooled-down
versions of RE\,0503$-$289 and, thus, PG\,1159 descendants. Note that both of
these WDs exhibit H as a trace element (i.e., they are DOAs) which might be an
indication that they went through a particular subclass of the various
final-thermal pulse (FTP) scenarios, the so-called AGB- (AFTP) and late-thermal
pulse (LTP) events (for details on these events see, e.g.,
\citealt{werner:06}). The lack of DAs with extreme trans-iron enrichments supports
the idea that a LTP is a necessary condition for the trans-iron overabundances in
the DO white dwarfs.

Future work should aim at the determination of all other trans-Fe elements
discovered in RE\,0503$-$289. Are they related to the s-process abundance
pattern or dominated by radiative levitation? For this, we need theoretical
predictions from diffusion calculations, in a manner pioneered by
\citet{chayer:2006} who performed such modeling for Ge, Zr, and Pb in sdB stars.
Such calculations should account for the effects of (selective) stellar
winds. Another ingredient for the diffusion calculations as well as for the
abundance analyses are oscillator strengths that must be obtained by quantum-mechanical calculations.

\section{Closing Remark}

In 1904, the Scottish chemist Sir William Ramsay received the chemistry Nobel prize
for the discovery of the noble gases He, Ne, Ar, Kr, and Xe in the air and for their
isolation. He earned his doctorate in 1873 -- at the University of T\"ubingen.

\acknowledgements T.R. is supported by DLR grant 05\,OR\,0806, and
E.R. by DFG grant WE1312/41-1.

\bibliography{werner}

\end{document}